\documentclass[12pt]{article}
\usepackage{theorem,amssymb,amsbsy,latexsym,amsmath,bbm,epsfig,psfrag}

\theoremstyle{plain}

{\theorembodyfont{\upshape}}

\begin{document}

\begin{center}

  {\Large \bf Matrix Models and Lorentz Invariance}

Jens Hoppe \\ 
\end{center}
\begin{center}
\scriptsize{Department of Mathematics, Royal Institute of Technology} \\
\scriptsize{100 44 Stockholm, Sweden}
\end{center}

\begin{center}
\noindent \textbf{Abstract} 
\end{center}
The question of Lorentz invariance in the membrane matrix model is addressed.\\
\section{Introduction}
It is generally believed that the matrix model Hamiltonian that was derived in [1] (concerning the bosonic theory) and [2],[3] (concerning the supersymmetric theory; see also [4]) is not Lorentz invariant (not even classically). When trying to either prove, or disprove, this belief, one finds quite a number of subtleties, and interesting connections, involving not only the (re)construction of $\zeta$ (conventionally called $x^-$) but also the fluid dynamics point of view [5], the no-go theorem for particle relativistic interactions [6], as well as sum-rules involving eigenvalues of (continuous and discrete) Laplacians, and dualities in the appearance of the Laplacian operators on the parameter-space, respectively embedded membrane.\\ One should mention that Goldstone [7] proved Lorentz invariance of the continuous (bosonic) theory in the mid-eighties (introducing some Green's function that allows to express $\zeta$ in terms of the transverse degrees of freedom), and that de Wit et al [8] - while extending Goldstone's analysis of the continuous theory to the supersymmetric one- came to the conclusion that the regularized theory 
is
not Lorentz invariant. \\ Two important new structures intimately related to Lorentz-invariance were found in [10].
\section{Canonical Light-Cone Description of M(em)branes}
Bosonic membranes can be described by ([1],[9])
\begin{equation}
H[\vec{x},\vec{p};\eta,\zeta_0]=\frac{1}{2\eta}\int\frac{\vec{p}^2+g}{\rho}d^2\varphi
\end{equation}
\begin{equation}  
        \int f^a \thinspace \vec{p} \cdot \partial_a \vec{x} \thinspace d^2\varphi = 0 \quad \textrm{whenever} \quad \partial_a(\rho f^a) = 0
\end{equation}
where
\begin{equation}
g=\det(\partial_a\vec{x}\cdot \partial_b\vec{x})
\end{equation}
and $\rho$ is a non-dynamical density of unit weight ($\int\rho(\varphi)d^2\varphi=1)$.
(1) and (2) can be thought of as arising from 
\begin{equation}
\widetilde{H}[\vec{x},\vec{p};\zeta, \pi]:=\int\frac{\vec{p}^2+g}{-2\pi}d^M\varphi
\end{equation}
\begin{equation}
C_a:=\pi\partial_a\zeta+\vec{p}\cdot\partial_a\vec{x}=0 \quad (a=1,...,M)
\end{equation}
via Hamiltonian reduction (using that $\dot{\pi}=-\frac{\delta\widetilde{H}}{\delta\zeta}=0$, cp.[1],[11]),
\begin{equation}
\pi(\varphi)=\rho(\varphi)\int\pi d^M\varphi= -\rho(\varphi)\eta, 
\end{equation} 
(and choosing M=2), resp. by comparison with the Lagrangian equations of motion
\begin{equation}
\frac{1}{\sqrt{G}}\partial_\alpha(\sqrt{G}G^{\alpha\beta}\partial_\beta x^{\mu})=0, \quad \mu=0,1,...,D-1
\end{equation}
in the gauge 
\begin{equation}
\varphi^0=\frac{x^0+x^{D-1}}{2}:=\tau, \quad G_{0a}=\partial_a\zeta-\dot{\vec{x}}\partial_a\vec{x}=0  ,
\end{equation}
in which (7), then implying
\begin{equation}
\frac{\partial}{\partial\tau}\sqrt{\frac{g}{2\dot{\zeta}-\dot{\vec{x}}^2}}=0 ,
\end{equation}
reduces to having to solve 
\begin{equation}
\Ddot{\vec{x}}=\frac{1}{\widetilde{\rho}}\partial_a(g\frac{g^{ab}}{\widetilde{\rho}}\partial_bx^{\mu}),
\end{equation}
(the equation for $\zeta:=x^0-x^{D-1}$ automatically follows), with $\widetilde{\rho}(\varphi) (\cong-\pi(\varphi))=\eta\rho(\varphi)$ being the time-independent density appearing in (9). $\zeta_0$ (mentioned in (1) as part of the canonical variables) is equal to the integral over $\zeta$ (a time-dependent constant that is not determined by (8)), and has the time-evolution (cp. (9))
\begin{equation}
\dot{\zeta_0}=-\frac{\delta H}{\delta\eta}=\frac{H}{\eta}.
\end{equation}
Finally, the mass-squared of the (internal degrees of freedom part of the) M(em)brane,
\begin{equation}
\mathbb{M}^2=\int\frac{\vec{p}^2+g}{\rho}-(\int \vec{p})^2=\sum_{i=1}^{D-2}\sum_{\alpha=1}^{\infty}p_{i\alpha}p_{i\alpha}+V
\end{equation}
\begin{equation*}
p_{i\alpha}:=\int p_iY_{\alpha}d^M\varphi, \quad x_{j\beta}:=\int x_jY_{\beta}\rho d^M\varphi 
\end{equation*}
\begin{equation*}
\int Y_{\alpha}Y_{\beta}\rho d^M\varphi=\delta_{\alpha \beta}, \quad \sum_{\alpha=1}^{\infty}Y_{\alpha}(\varphi)Y_{\alpha}(\widetilde{\varphi)}=\frac{\delta(\varphi, \widetilde{\varphi})}{\rho(\varphi)}-1 
\end{equation*}
then takes a very simple form (cp.[1]), with the interaction $V$ determined entirely in terms of the structure constants of the Lie-algebra of diffeomorphisms, $\varphi^a\rightarrow \widetilde{\varphi}^a$, with unit Jacobian (which is the residual diffeomorphism symmetry group left after gauge fixing). 
$\zeta$ (seemingly having disappeared) is needed both geometrically (to reconstruct the world volume swept out by the membrane in space-time) and for the Poincare'-invariance of (12) ( as well as returns as the central object in the fluid dynamic description, after a hodograph transformation [5] ) Obviously, $\mathbb{M}^2$ commutes with $H$ and $\vec{P}=\int \vec{p}d^M\varphi=\vec{p_0}$ ( which generate light-cone time- and space translations), as well as with $\vec{P}_+=\eta=-\int \pi d^M\varphi$, $\zeta_0\eta$, and $\eta\int x_i\rho d^M\varphi$, while 
\begin{equation}
M_{i-}=\int(x_i\frac{\vec{p}^2+g}{2\eta\rho}-\zeta p_i)d^M\varphi
\end{equation}
necessitates the reconstruction of $\zeta$ (in terms of the other degrees of freedom) from (cp. (8), (9))
\begin{equation}
\eta\partial_a\zeta=\frac{\vec{p}}{\rho}\partial_a\vec{x}
\end{equation}
\begin{equation}
\eta^2\dot{\zeta}=\frac{1}{2}\frac{\vec{p}^2+g}{\rho^2},
\end{equation}
consistent provided that the equations of motion (cp. (10)/(1))
\begin{equation}
\eta\dot{\vec{x}}=\frac{\vec{p}}{\rho},\quad \eta\frac{\dot{\vec{p}}}{\rho}=\frac{1}{\rho}\partial_a(g\frac{g^{ab}}{\rho}\partial_b\vec{x})
:=\mathbf{\Delta}\vec{x},
\end{equation}
( and the constraint 
(2)) hold.
\section{Reconstructing $\zeta$}
The reconstruction of $\zeta$ from
\begin{equation*}
\eta\partial_a\zeta=\frac{\vec{p}}{\rho}\partial_a\vec{x}, \quad 2\eta^2\dot{\zeta}=\frac{\vec{p}^2+g}{\rho^2}\quad (*)
\end{equation*}
is [7]
\begin{equation}
\overline{\zeta}:=\eta(\zeta-\zeta_0)=\int G(\varphi,\widetilde{\varphi})\widetilde{\nabla}^a(\frac{\vec{p}}{\rho}\widetilde{\partial}_a\vec{x})\rho d^M\widetilde{\varphi}
\end{equation}
where
\begin{equation}
G(\varphi,\widetilde{\varphi}):=\sum_{\alpha=1}^{\infty}\frac{-1}{\mu_{\alpha}}Y_{\alpha}(\varphi)Y_{\alpha}(\widetilde{\varphi})
\end{equation}
(conveniently choosing the $Y_{\alpha}$ as eigenfunctions of $\Delta$, $\Delta Y_{\alpha}=-\mu_{\alpha}Y_{\alpha}$) 
satisfies
\begin{equation}
\Delta G(\varphi,\widetilde{\varphi})=\frac{\delta(\varphi, \widetilde{\varphi})}{\rho(\varphi)}-1, 
\end{equation}
and $\vec{x}$ and $\vec{p}$ must satisfy
\begin{equation}
   \int f^a \thinspace \vec{p} \cdot \partial_a \vec{x} \thinspace d^M\varphi = 0 \quad \textrm{whenever} \quad \nabla_a f^a = 0  .
\end{equation}
To rewrite (17) in a form that suggests a canonical discrete analogue of $\zeta$
one just needs to note that 
\begin{equation}
\nabla^au\nabla_av=\frac{1}{2}(\Delta(uv)-u\Delta v-v\Delta u)  ;
\end{equation}
so (17) can be written as (note the 3 typos in eq.(7) of [10], and one missing ${\rho}$ in eq.(5) of [12])
\begin{equation}
2\overline{\zeta}=\frac{\vec{p}\cdot\vec{x}}{\rho} -\int\frac{\vec{p}\cdot\vec{x}}{\rho}+ \int G(\varphi,\widetilde{\varphi})(\frac{\vec{p}}{\rho}\Delta\vec{x}-\vec{x}\Delta\frac{\vec{p}}{\rho})\rho d^M\widetilde{\varphi}
\end{equation}
from which it follows that (cp.(16))
\begin{equation}
2\eta\dot{\overline{\zeta}}=\frac{\vec{p}^2}{\rho^2}-\int\frac{\vec{p}^2}{\rho^2}+ \int G(\varphi,\widetilde{\varphi})(\mathbf{\Delta}\vec{x}\Delta\vec{x}-\mathbf{\Delta}\vec{x}\Delta\vec{x})\rho d^M\widetilde{\varphi} + \vec{x}\mathbf{\Delta}\vec{x}-\int\vec{x}\mathbf{\Delta}\vec{x}.
\end{equation}
(implying that the terms not involving $\vec{p}$ give $g/\rho^2-\int g/\rho^2$).\\
Inserting (18) and using that $\Delta\vec{x}=-\mu_{\beta}\vec{x}_{\beta}Y_{\beta}$ etc. one finds that
\begin{equation}
2\overline{\zeta}=\sum_{\alpha,\beta,\gamma}\frac{\mu_{\alpha}+\mu_{\beta}-\mu_{\gamma}}{\mu_{\alpha}}\vec{p}_{\gamma} \cdot \vec{x}_{\beta}d_{\alpha \beta \gamma}Y_{\alpha}(\varphi)+2\vec{P}(\vec{x}-\vec{X}) ,
\end{equation}
implying e.g. that (for general M, \{...\} being totally antisymmetric, and linear in the derivatives of the M entries)
\begin{equation}
\eta\sum_{\alpha,\beta}d_{\alpha \beta \gamma}\frac{\mu_{\alpha}+\mu_{\beta}-\mu_{\gamma}}{\mu_{\alpha}}\dot{\vec{p}}_{\gamma} \cdot \vec{x}_{\beta}=\frac{1}{M!}d_{\alpha \beta \gamma}\{x_{i_1},...,x_{i_M}\}_{\beta}\{x_{i_1},...,x_{i_M}\}_{\gamma}  ,
\end{equation}
rather nontrivial identities involving the eigenvalues of the Laplacian $\Delta$ and the totally (anti-)symmetric structure constants
\begin{equation}
d_{\alpha \beta \gamma}:=\int Y_{\alpha}Y_{\beta}Y_{\gamma}\rho d^M\varphi , \quad   g_{\alpha \alpha_1 ...\alpha_M}:=\int Y_{\alpha}\{Y_{\alpha_1},...,Y_{\alpha_M}\}\rho d^M\varphi;
\end{equation}
note that (16) can be written in terms of the multilinear brackets (cp.[1]) as
\begin{equation}
\eta \frac{\dot{\vec{p}}}{\rho}=\frac{1}{(M-1)!}\{x_{i_1},...,x_{i_{M-1}},\{x_{i_1},...,x_{i_{M-1}},\vec{x}\}\}.
\end{equation}
\section{Matrix Analogues}
One possibility to define (for M=2) a matrix $\zeta_N$, that could then be used for 
\begin{equation}
2\eta M_{i-}=\text{Tr}(X_i(\vec{P}^2+W)-2\eta\zeta_{N}P_i)
\end{equation}
would be to use (24), i.e. define
\begin{equation}
2\overline{\zeta}_N:=\sum_{1}^{N^2-1}\frac{\mu_{a}+\mu_{b}-\mu_{c}}{\mu_{a}}\vec{p}_c\cdot\vec{x}_bd_{abc}^{(N)}T_a^{(N)}+2\vec{p}\cdot(\vec{X}-\vec{x}) .
\end{equation}
(denoting the transverse zero-modes now by x and p, and the matrices by X and P).
While it is convenient to assume
\begin{equation}
\frac{1}{N}\text{Tr}(T_a^{(N)}T_b^{(N)})=\delta_{ab}, \quad {\Delta}_N T_a^{(N)}=-\mu_a^{(N)}T_a^{(N)}
\end{equation}
\begin{equation*}
d_{abc}^{(N)}:=\frac{1}{2N}\text{Tr}(\{T_a,T_b\}T_c), \quad {\Delta}_N=\frac{-1}{\hbar^2}[T_n,[T_n,\cdot]],
\end{equation*}
corresponding to the matrix regularization
\begin{equation}
Y_{\alpha}(\varphi)\rightarrow T_a^{(N)}.
\end{equation}
it is in practice often simpler to have complex eigenfunctions $Y_{\alpha}$ and non-hermitean $T_a^{(N)}$, (for the sphere, e.g., see [1], [13]; $a=(lm)$ and $\mu_{l<N,m}^{(N)}=\mu_{lm}$, independent of $N$, $\hbar=\frac{2}{\sqrt{N^2-1}}$, $T_{l\geq N, m}^{(N)}\equiv0$ ). \\ The problem with (29) (which looks like a nice definition) is that it is degenerate for $N=2$ (where also the invariance requirement that the dynamical Poisson-bracket of $L_{abc}\vec{x}_b\vec{p}_c$ and 
$\epsilon_{a'b'c'}\vec{x}_{b'}\vec{p}_{c'}$ should equal $\epsilon_{aa'e}L_{edf}\vec{x}_d\vec{p}_f$ on the constrained phase-space does not have any non-trivial solution) , while for higher N difficult to test.

Already in the (much simpler) continuous case, it is quite non-trivial (even for the string, and torodial membranes) to verify (24) directly, i.e. without going back to (17). Differentiating (17), on the other hand, with respect to (light-cone) time, one trivially gets $\vec{p}^2/(2\eta\rho^2)-\int(\vec{p}^2/(2\eta\rho^2))$ and due to 
\begin{equation}
2\nabla^a(\frac{1}{\rho}\partial_b(g\frac{g^{bc}}{\rho}\partial_c\vec{x})\partial_a\vec{x})=\nabla^a(\frac{2}{\rho}\partial_b(\delta_a^bg/\rho)-\frac{1}{\rho^2}\partial_ag)=\Delta(g/\rho^2)
\end{equation}
\begin{equation}
\eta\int G(\varphi,\widetilde{\varphi})\widetilde{\nabla}^a(\dot{\vec{p}}/\rho\widetilde{\partial}_a\vec{x})d^M\widetilde{\varphi}=\frac{g}{2\rho^2}-\int\frac{g}{2\rho^2} 
= -\int\nabla^aG\partial_a\vec{x}\mathbf{\Delta}\vec{x}
\end{equation}
(indirectly proving (25)). When trying to use eq.(27)
for the proof of (33), one can make use of an identity, that for M=2 reads
\begin{equation}
\{x_j,\{x_j,\vec{x}\}\}\{\vec{x},Y_{\beta}\}=\frac{1}{4}\{\{x_i,x_j\}^2,Y_{\beta}\} \quad \forall \beta
\end{equation}
and whose discrete analogue,
\begin{equation}
[X_j,[X_j,\vec{X}]][\vec{X},T_b] =\frac{1}{4}[[X_i,X_j]^2,T_b]
\end{equation}
does not hold in general - but would need to hold at least for \emph{some} $T_b$ (those entering the Laplacian on the discrete base-manifold).
While this makes discrete Poincare' invariance rather unlikely, it does not prove that there is not perhaps another way to define a matrix $\zeta_N$ for which (restricting for the moment to just finding quantities constant in time)
\begin{equation}
\text{Tr}(\eta X_i\dot{\zeta}_N-\zeta_NP_i).
\end{equation}
Using the discrete equations of motion, 
\begin{equation}
\eta\dot{X}_i=P_i, \quad \eta\dot{P}_i=-[X_j,[X_j,X_i]]:=\mathbf{\Delta}_NX_i
\end{equation}
one finds that the only requirement for (36) to be time-independent is that $\zeta_N$, just as $\vec{X}$ (hence: as it should), satisfy the discrete analogue of what follows from (14) and (15) (cp. (10)),
\begin{equation}
\eta^2\Ddot{\zeta}=\mathbf{\Delta}_N\zeta.
\end{equation}
A solution of (38), in terms of $\vec{X}$ and $\vec{P}$ satisfying (37) and the usual "Gauss-law" matrix model constraint 
\begin{equation}
[\vec{X},\vec{P}]=0 \quad ,
\end{equation}
can be constructed when trying to discretize (14) by noting that in the continuous case one certainly has
\begin{eqnarray}
\eta\{\zeta,x_i\} &=& \frac{\vec{p}}{\rho}\{\vec{x},x_i\}\\
\eta\{\zeta,p_i\} &=& \frac{\vec{p}}{\rho}\{\vec{x},p_i\}
\end{eqnarray}
-- whose canonical discrete analogues would be 
\begin{eqnarray}
2\eta[\zeta_N,X_i]&=&\vec{P}[\vec{X},X_i]+[\vec{X},X_i]\vec{P} \\
2\eta[\zeta_N,P_i]&=&\vec{P}[\vec{X},P_i]+[\vec{X},P_i]\vec{P}.
\end{eqnarray}
Using (42), (43) and (39), it easily follows that $\zeta_N$ does satisfy (38), provided 
\begin{equation}
2\eta\dot{\zeta}_N=\vec{P}^2-\frac{1}{2}[X_j,X_k]^2 \quad ,
\end{equation} 
whose commutator with $X_i$ is consistent with what one gets from differentiating (42), using (43). The problem with this solution however is that in contrast to the continuous case, (42) and (43) (even one of the two by itself) are not necessarily consistent, i.e.,
for consistency require further identities to be satisfied by the $X_i(\tau)$ and $P_j(\tau)$.\\
E.g., commuting (42) with $P_i$, and subtracting (43) commuted with $X_i$ (then summing over i) one obtains (using (39))
\begin{equation}
2[P_i,X_j][X_i,P_j]=[P_i,P_j][X_i,X_j]+[X_i,X_j][P_i,P_j].
\end{equation} 
The difference between discrete and continuous cases becomes most drastic by noting that
\begin{equation}
\eta[2\zeta_N,T_a]=\vec{P}[\vec{X},T_a]+[\vec{X},T_a]\vec{P}
\end{equation} 
- in a way, the most natural generalization of  
\begin{equation}
\eta\partial_a\zeta=\frac{\vec{p}}{\rho}\partial_a\vec{x}, \quad \text{resp.} \quad \eta\{\zeta,Y_{\alpha}\}=\vec{p}\{\vec{x},y_{\alpha}\}, 
\end{equation} 
by commutating again with $T_a$ (and noting that $[T_a,[T_a,\cdot]]$ is the adjoint Casimir, $\chi_N$
) would actually imply that 
\begin{equation}
2\eta \zeta_N=\vec{P} \cdot \vec{X}+ \vec{X} \cdot \vec{P}+\frac{1}{\chi_N}([\vec{P},T_a][\vec{X},T_a]+[\vec{X},T_a][\vec{P},T_a])
\end{equation} 
would be symmetric under $\vec{X}\leftrightarrow \vec{P}$, in sharp contrast to (22). Note that in the continuous case a similar argument is prevented by the value of the corresponding "Casimir" (for the infinite -dimensional Lie algebra) being infinite. What one \emph{could} do, however, is to demand (46) only for \emph{some} $T_a$, namely those entering the Laplacian of the discretized base manifold;
for the 2-sphere e.g. (cp. [1],[13]) these would be the 3 generators $T_{l=1,m=-1,0,+1}$, thus obtaining 
\begin{equation}
2\eta\Delta_N\zeta_N=\vec{P}\Delta_N\vec{X}+\Delta_N\vec{X}\vec{P}-\frac{1}{\hbar^2}([T_m,\vec{P}][T_m,\vec{X}]+[T_m,\vec{X}][T_m,\vec{P}]).
\end{equation} 
This approach also leads to (29). \\ Finally, one could try to directly  solve
\begin{equation}
2\eta\dot{\zeta_N}=\frac{\vec{P}^2+W}{\eta}, \quad W=-\frac{1}{2}[X_i,X_j]^2
\end{equation} 
but one should be aware of possible renormalizations.\\
In any case it would be interesting to see a critical dimension when quantizing.\begin{center}
\noindent \textbf{Acknowledgement} 
\end{center}
I would like to thank the students of my KTH course ``Dynamics of Strings and Membranes'' (which this note, written in March 2010, is based on) for their interest.
\begin{center}
\noindent \textbf{Note Added} 
\end{center}
The original equations determining $\zeta$ (cp.(14),(15)) are invariant under diffeomorphisms having unit Jacobian. Therefore (22)/(24) should, on the constrained phase space, also have this property, despite the dependence of $\Delta$ (hence its eigenvalues) on the metric chosen to obtain $\zeta$ explicitely. Indeed, one can check by an explicit (somewhat non-trivial)
calculation (for simplicity I chose M=2) that the dynamical Poisson-bracket of $\zeta_{\alpha}$ with 
$g_{\beta\gamma\epsilon}\vec{x}_{\gamma}\vec{p}_{\epsilon}$ , i.e. the $\beta$-component of $\phi:=\{\frac{\vec{p}}{\rho},\vec{x}\}=\frac{\epsilon^{ab}}{\rho}\partial_a(\frac{\vec{p}}{\rho})\partial_b\vec{x}$, is $g_{\alpha\beta\gamma}\zeta_{\gamma}$ on the constrained phase space (implying in particular that $M_{i-}$ is well defined, i.e. commutes with the constraint $\phi$).  \quad
I thank Joakim Arnlind, Martin Bordemann, Ki-Myeong Lee and Piljin Yi for discussions related to this point.

\begin{center}
\noindent \textbf{References} 
\end{center}
$[1]$ Jens Hoppe \textit{Quantum Theory of a Massless Relativistic Surface and ...}
Ph.D. Thesis, MIT 1982, {\tt http://dspace.mit.edu/handle/1721.1/15717 }
; 
see also \textit{Membranes and Matrix Models},
{\tt arXiv:hep-th/0206192}, IHES/P/02/47, and references therein.\\ 
$[2]$ M.Claudson,M.B.Halpern; Nucl.Phys.B250 (1985) 689.\quad
    R.Flume, Ann.Phys.164 (1985) 189.\quad
    M.Baake, P.Reinicke, V.Rittenberg; J.Math.Phys.26 (1985) 1070 \\
$[3]$ B.deWit, J.Hoppe, H.Nicolai, Nucl.Phys.B305 (1989) 545.\\
$[4]$ T.Banks, W.Fischler, S.Shenker, L.Susskind, Phys.Rev.D55 (1997) 6189\\
$[5]$ M.Bordemann, J.Hoppe; Phys.Lett.B325 (1994) 359\\
$[6]$  D.G. Currie, T.F. Jordan, E.C.G. Sudarshan; Rev. Mod. Phys. 35
    (1965) 350.\\
    H. Leutwyler; Nuovo Cim. 37 (1965) 556.\\
$[7]$ J. Goldstone, unpublished notes (
passed on to the authors of [8] in 1987/88 )\\
$[8]$ B.deWit, U.Marquard, H.Nicolai, Comm.Math.Phys.128 (1990) 39\\
$[9]$ J.Goldstone; unpublished\\
$[10]$ J.Hoppe, arXiv:1003.5189\\
$[11]$ J.deWoul, J.Hoppe, D.Lundholm; arXive:1006.4714\\
$[12]$ J.Hoppe; ``Quantum Reconstruction Algebras''\\ 
http://urn.kb.se/resolve?urn=urn:nbn:se:kth:diva-13591\\
$[13]$ J.Hoppe, S.T.Yau; Comm.Math.Phys.195 (1998) 67.

\end{document}